\documentclass[twocolumn,preprintnumbers,nofootinbib,superscriptaddress]{revtex4}
\usepackage{amsmath} \usepackage{graphicx} \usepackage{amsfonts}
\usepackage{array} \usepackage{amsthm} \usepackage{bm} %\usepackage{mathptmx}

\usepackage{latexsym}

\usepackage[breaklinks]{hyperref}
\usepackage{color}

\newcommand{\be}{\begin{equation}}
\newcommand{\ee}{\end{equation}}
\newcommand{\ba}{\begin{eqnarray}}
\newcommand{\ea}{\end{eqnarray}}
\newcommand{\bal}{\begin{align}}
\newcommand{\eal}{\end{align}}

\newcommand{\e}{{\rm e}}
\newcommand{\dd}{{\rm d}}

\newcommand{\bb}{\bibitem}

\newcommand{\la}{\lambda}

\newcommand{\bt}{\beta}

\newcommand{\ga}{\gamma}

\newcommand{\Si}{\Sigma}

\newcommand{\Om}{\Omega}

\newcommand{\de}{\delta}

\newcommand{\bw}{\begin{widetext}}
\newcommand{\ew}{\end{widetext}}

\def\abh{black hole }
\def\bh{black holes }

\def\BH{black holes}
\def\RN{Reissner-Nordstr\"om }
\def\RNNB{Reissner-Nordstr\"om}

\def\tv{thermodynamic variables }
\def\TV{thermodynamic variables}
\def\tp{thermodynamic potential }

\begin{document}
%\begin{flushleft}???
%\end{flushleft}
%\begin{flushleft}???
%\end{flushleft}
\title{Geometrothermodynamics: Comments, critics, and supports}

\author{Mustapha Azreg-A\"{\i}nou}%\email{azreg@baskent.edu.tr}
\affiliation{Ba\c{s}kent University, Engineering Faculty, Ba\u{g}l\i ca Campus, Ankara, Turkey}

%\date{}

\begin{abstract}
We write explicitly the Euler identity and the Gibbs-Duhem relation for thermodynamic potentials that are not homogeneous first-order functions of their natural extensive variables. We apply the rules to the theory of geometrothermodynamics and show how the use of the natural extensive variables, instead of the modified ones, leads to misleading results. We further reveal some other ambiguities and inconsistencies in the theory and we make new suggestions.
\end{abstract}

%\pacs{04.20.-q, 04.70.Dy, 02.40.Ky}

\maketitle

\section{Introduction}

There is a couple of theories on the geometry of thermodynamics which have been applied to \abh thermodynamics~\cite{W1,W2,R,LLLS,hd}. The metrics by Weinhold~\cite{W1,W2} and Ruppeiner~\cite{R} have received critics for not being Legendre invariant~\cite{gbh}. For the Ruppeiner metric, however, this shortcoming  has been remedied by proving the existence of a one-to-one correspondence between the divergencies of the heat capacities and those of the curvature scalars for thermodynamic descriptions where the potentials are related to the mass (instead of the entropy) by Legendre transformations~\cite{cor}. This has resulted in a full agreement of the classical and the geometric descriptions of the \abh thermodynamics for most of applications met in the literature~\cite{cor} and thus has corroborated the theory of the geometry of thermodynamics. While the theory by Liu, L\"{u}, Luo and Shao~\cite{LLLS} has only received supports so far~\cite{S,O1}, the geometrothermodynamics (GTD) by Quevedo~\cite{gbh} has been subject to both critics~\cite{M1,M2} and supports~\cite{S,O1} from a physical point of view. This work presents a first critic to GTD from a mathematical as well as a physical point of view.

Prior to this critic, we have generalized in~\cite{conf2} the change of representation formula derived mostly for GTD application purposes by Quevedo et al.~\cite{conf}. Such generalizations allow us to include all physical applications, particularly, applications to black hole thermodynamics, cosmology and fluid thermodynamics.

Since this work is a series of comments and critics on GTD, more precisely on the conclusions derived by GTD, we assume that this theory is known to readers and refer them to the work by Quevedo et al.~\cite{hd,gbh}.

The remaining part of this work is divided into two sections and an appendix. In Sect.~\ref{sec2} we introduce two types of extensive \TV, the natural ones, $E^a$, are used to express the first law of thermodynamics and the modified variables, $E^{\,\prime a}$, in terms of which the \tp is a homogeneous function of some order, say, $\bt$.

The use of $E^a$, instead of the modified extensive variables $E^{\,\prime a}$, can lead to misleading results in GTD and any other fields~\cite{d2,new,newa,newb} where potentials which are not homogeneous first-order functions are used. We particularly show how the confusion of these sets of extensive \tv was the source of misleading conclusions and derivations by the authors of GTD. We will also derive a generalized Euler identity, that is, an Euler identity for thermodynamic potentials that are not homogeneous first-order functions, as well as a generalized Gibbs-Duhem relation applicable to a wide range a physical problems and other useful relations. These derivations do not constitute the main purpose of this work; rather, they constitute a tool for revealing discrepancies of GTD and suggesting possible remedies.

In \abh thermodynamics the use of the modified extensive variables $E^{\,\prime a}$ was first introduced by Davies~\cite{d2}. Further developments have led to the formulation of the postulates of gravitational thermodynamics~\cite{new2} where it was clearly emphasized that ``Fundamental equations are in general no longer homogeneous first-order functions of their extensive variables". The analysis developed in Sect.~\ref{sec2}, concerning the introduction of the modified extensive variables $E^{\,\prime a}$, follows closely that made in~\cite{d2}.

In Sect.~\ref{sec3} we comment on a series of papers by Quevedo et al. In the appendix, we derive a useful relation, that is the Smarr formula for Kerr \abh in $d$-dimensions, needed in Sect.~\ref{sec3}.

Our main purpose in commenting on GTD and criticizing it is to provide a platform for improving the theory, which has received supports from other workers as mentioned earlier in this section. In Sect.~\ref{sec4} we draw our conclusions concerning possible remedies to the theory.

\section{Homogenous potentials \label{sec2}}

In this work we use the convention by which repeated indices are summed except when otherwise mentioned. We use the same notations as in~\cite{hd} to denote the thermodynamic quantities. This has always been the same notation in all papers on GTD. Hence, ($E^a,I^a$) denote extensive and intensive thermodynamic variables, respectively, with $I_a(E^a)=\partial\Phi/\partial E^a$  ($I_a=\de_{ab}I^b$) and $\Phi(E^a)$ is some thermodynamic potential. The first law of thermodynamics takes the form
\begin{equation}\label{1}
    \dd\Phi=I_a\,\dd E^a\quad (\Si \text{ over }a,\;a=1,2,\ldots ).
\end{equation}

The knowledge of $\Phi$ is crucial for the determination of the thermodynamic properties of the system under consideration and for its phase transitions. In classical thermodynamics, $\Phi$ is a homogeneous first-order function of the variables $E^a$, which are called the natural variables~\cite{new3}, and the $I^a$'s are homogeneous zero-order functions of their extensive variables. Equations $I_a(E^a)\equiv \partial\Phi/\partial E^a$ are called equations of state.

In some thermodynamical problems~\cite{d2,new,newa,newb}, including \BH, $\Phi$ appears to be homogeneous of some other set of extensive variables~\cite{conf2}, denoted here by $E^{\,\prime a}$, which is in general different from the natural set $E^a$ in terms of which the first law~\eqref{1} is formulated (as we shall see below, there are cases where $\Phi$ is not homogeneous at all). This is to say that in some fields of thermodynamics, $\Phi$ is not homogeneous first-order function of its natural extensive variables $E^a$, contrary to one of the postulates of classical thermodynamics.

To our knowledge, in all cases of interest, particularly in \bh thermodynamics as we shall see below, the variables $E^{\,\prime a}$ are power-law functions of $E^a$:
\begin{equation}\label{n1}
E^{\,\prime a}=(E^a)^{p_a}\quad \text{(no summation over } a),
\end{equation}
where $p_a$ depends obviously on $a$. It was shown in~\cite{conf2} that $p_a$ depends also on $\bt$:
\begin{equation}\label{1b}
    p_a\equiv p_a(\bt),
\end{equation}
where $\bt$ is the order of homogeneity of $\Phi$. We shall re-derive~\eqref{1b} in this section and show that we can always choose $\bt=1$. In the case of~\eqref{n1}, this means that we can always make $\Phi$ homogeneous first-order function of the modified extensive variables $(E^a)^{p_a}$ instead of the natural ones $E^a$.

Before we give some examples from \abh thermodynamics, we first consider the generic case where $\Phi$ is homogeneous in $E^{\,\prime a}$ of order $\bt$: $\Phi(\la E^{\,\prime a})=\la^{\bt}\Phi(E^{\,\prime a})$. We restrict ourselves to the case of interest~\eqref{n1}, then by Euler theorem we obtain
\begin{align}
\label{2}\bt\Phi & = E^{\,\prime a}\frac{\partial\Phi}{\partial E^{\,\prime a}}\quad (\Si \text{ over }a)\\
\label{3}\quad & = \frac{E^a}{p_a}\frac{\partial\Phi}{\partial E^a}\quad (\Si \text{ over }a)
\end{align}
where we have used $\partial E^{\,\prime a}/\partial E^a=p_a(E^a)^{p_a-1}$ (no summation over $a$). Eq.~\eqref{3} generalizes Euler identity to cases where the potential $\Phi$ fails to be homogeneous in the natural \tv $E^a$ in terms of which the first law~\eqref{1} is formulated. Thus, in general, we have
\begin{equation}\label{3b}
    \bt\Phi\neq E^a\partial\Phi/\partial E^a.
\end{equation}

We have noticed that the authors of GTD, Quevedo et al., have always assumed $\bt\Phi\equiv E^a\partial\Phi/\partial E^a$ (or resp. $\Phi\propto E^a\partial\Phi/\partial E^a$), thus they have admitted that all $p_a\equiv 1$ (or resp. all\footnote{When all $p_a$ are equal, it is safe to write $\Phi\propto E^a\partial\Phi/\partial E^a$ but it is neither correct nor is it safe, as we shall see in case (c) of Sect.~\ref{sec3} concerning Kerr \bh in $d$-dimensions, to assume and use the equality $\Phi = E^a\partial\Phi/\partial E^a$.} $p_a$ are equal), which is, from the one hand, a very restrictive constraint and rarely met in \abh thermodynamics, cosmology, fluid thermodynamics or other fields of thermodynamics and, from the other hand, the constraint was applied indiscriminately to all problems the authors have tackled even when $\Phi$ was not homogeneous at all! We have realized that their assumption occurred in the paragraph following Eq. (37) of Ref.~\cite{hd}, in Eqs. (2), (4) and (11) [and probably (12)] of Ref.~\cite{gbh}, in the paragraph following Eq. (13) of Ref.~\cite{gbh}, in Eq. (4) of Ref.~\cite{ft}, in the paragraph following Eq. (6) of Ref.~\cite{ft}, in the paragraph following Eq. (33) of Ref.~\cite{btz}, and in Eq. (6) of Ref.~\cite{gt}; it has occurred in other related papers too as we shall see below and recently in Eq. (1) of~\cite{rep}.

Before we proceed further with Eqs.~\eqref{2} and~\eqref{3}. We first give an example from black hole thermodynamics. Some other examples are provided in~\cite{O1,conf2,d2,O2,O3}. Consider the \RN \abh where its mass is taken as a thermodynamic potential~\cite{d} (see also~\cite{gbh})
\begin{equation}\label{4}
    M=(\pi S^{-1/2}Q^2+S^{1/2})/(2\sqrt{\pi}).
\end{equation}
The natural extensive \tv that enter the first law are ($S,Q$):
\begin{equation}\label{5}
    \dd M=T\dd S+\phi \dd Q
\end{equation}
where
\begin{equation}\label{n2}
T=(\partial M/\partial S)_Q,\quad \phi=(\partial M/\partial Q)_S
\end{equation}
are the temperature and electric potential given by
\begin{equation}\label{5b}
T=S^{-3/2}[S-\pi Q^2]/(4\sqrt{\pi}),\;\phi=\sqrt{\pi}S^{-1/2}Q.
\end{equation}

Now, it is straightforward to check that $M$ is not homogeneous in ($S,Q$) because it is not possible to find a real $\bt$ such that $M(\la S, \la Q)=\la^{\bt}M(S,Q)$; rather, it is homogeneous in ($S,Q^2$) of order $\bt=1/2$
\begin{equation}\label{6}
   M(\la S, \la Q^2)=\la^{\bt}M(S,Q^2) \quad \text{with }\bt=1/2
\end{equation}
leading to the Euler identity~\eqref{2}, \eqref{3}
\begin{align}
\label{7}M/2&=S(\partial M/\partial S)_Q+Q^2[\partial M/\partial (Q^2)]_S\\
\label{7b}\quad &=ST+Q\phi/2
\end{align}
where we have used the definitions~\eqref{n2} of $T$ and $\phi$ along with, $[\partial M/\partial (Q^2)]_S=[\partial M/\partial Q]_S/(2Q)$, $p_1\equiv p_S=1$ and $p_2\equiv p_Q=2$. It is straightforward to check that the right-hand side of~\eqref{7b} is equal $M/2$ on substituting the expressions of $T$ and $\phi$ given in~\eqref{5b}.

Now, rewriting the expression~\eqref{4} of $M$ as:
\begin{equation}
M=[\pi (S^{\ga})^{-1/(2\ga)}(Q^{2\ga})^{1/\ga}+(S^{\ga})^{1/(2\ga)}]/(2\sqrt{\pi}),
\end{equation}
where $\ga >0$, one sees that the same function $M$ is also homogeneous in ($S^{\ga},Q^{2\ga}$) of order $\bt=(1/2)/\ga$. For instance, if we choose $\ga =3$, leading to $p_1\equiv p_S=\ga=3$ and $p_2\equiv p_Q=2\ga=6$, we obtain using~\eqref{3}
\begin{equation}\label{note}
\frac{M}{6}=S^3\Big(\frac{\partial M}{\partial (S^3)}\Big)_Q+Q^6\Big(\frac{\partial M}{\partial (Q^6)}\Big)_S=\frac{ST}{3}+\frac{Q\phi}{6},
\end{equation}
which is identical to~\eqref{7b}. If one chooses $\ga=1/2$, the same expression~\eqref{4} of $M$ appears homogeneous in ($S^{1/2},Q$) of order $\bt=1$ with $p_1\equiv p_S=1/2$ and $p_2\equiv p_Q=1$. As one sees, there is a one-to-one correspondence:
\begin{equation}
 \text{order of homogeneity } \leftrightarrow \text{ values of } p_a\text{'s}.
\end{equation}

As a general rule: if $f$ is homogeneous in ($x,y,\ldots$) of order $\bt$ then it is also homogeneous in ($x^{\ga},y^{\ga},\ldots$) of order $\bt/\ga$. Since $\ga$ is arbitrary, this means that the order of homogeneity can be any number one chooses, a one particular choice is $\ga=\bt$ by which $f$ is rendered homogeneous in ($x^{\bt},y^{\bt},\ldots$) of order 1. This means that one can always fix the value of the order of homogeneity to 1~\cite{conf2} by modifying the values of the powers $p_a$, which depend on the order of homogeneity as we have seen in our previous example, and conversely, the order of homogeneity depends on $p_a$'s.

If, now, $\bt$ is some generic order of homogeneity of $\Phi$, it is clear that~\eqref{1b} holds.

Now back to~\eqref{3}. On dividing both sides of this equation by $\bt$ we obtain
\begin{equation}\label{3c}
   \Phi= \frac{E^a}{\bar{p}_a}\frac{\partial\Phi}{\partial E^a}\quad (\Si \text{ over }a)
\end{equation}
where $\bar{p}_a\equiv \bt p_a(\bt)$. Here $\Phi$ appears as homogenous in $(E^a)^{\bar{p}_a}$ of order 1. Thus, the powers $\bar{p}_a$ are those associated with an order of homogeneity equal 1. The importance of $\bar{p}_a$ is that they depend neither on a particular choice of the order of homogeneity nor on the values of $p_a$'s. If a generic value $\bt$ of the order of homogeneity is known along with $p_a$'s, as in the previous example, then
\begin{equation}\label{3d}
    \bar{p}_a = \bt p_a(\bt),
\end{equation}
where the right-hand side does not depend on a particular choice of $\bt$, as this can easily be checked using the different values of the order of homogeneity in the example of the function $M$ given by~\eqref{4}.

Another useful generalization is that of the Gibbs-Duhem relation, which on using~\eqref{3}, takes the form
\begin{equation}
    \frac{E^a}{p_a}\,\dd I_a=\Big(\bt -\frac{1}{p_a}\Big)I_a\dd E^a \quad (\Si \text{ over }a),
\end{equation}
or, equivalently, the form
\begin{equation}\label{n3}
    \frac{E^a}{\bar{p}_a}\,\dd I_a=\Big(1 -\frac{1}{\bar{p}_a}\Big)I_a\dd E^a \quad (\Si \text{ over }a).
\end{equation}
One sees that only in the case where all $\bar{p}_a\equiv 1$, the relation~\eqref{n3} reduces to the classical-thermodynamic Gibbs-Duhem one: $E^a\dd I_a=0$. In the case where all $\bar{p}_a$ are equal but different from 1, Eq.~\eqref{n3} is still different from, and generalizes, the classical-thermodynamic Gibbs-Duhem relation.

\section{Comments and critics \label{sec3}}

We now see some of the consequences of the above-mentioned assumption and give our first example of misleading results in GTD where Quevedo et al. assumed that $E^a\partial\Phi/\partial E^a$ is proportional to $\Phi$ when, according to~\eqref{3b} or\eqref{3c}, it is not.

\vspace{3mm}

\paragraph*{\textbf{(a) \RN \bh in $d$-dimensions.}} Consider Eq. (20) of~\cite{hd}, which we rewrite setting
\begin{equation}\label{g0}
D\equiv (d-3)/(d-2)
\end{equation}
as
\begin{equation}\label{q1}
    H(S,\phi)=-S^D(2D\phi^2-1)/2=-S^D\mathcal{B}_3/[2(d-2)]
\end{equation}
where the correct expression of $\phi$ is:
\begin{equation}\label{q1b}
\phi=Q/(2DS^D)
\end{equation}
instead of $\phi=Q/(2DS^{1/D})$ as given in Eq. (13) of~\cite{hd} and the temperature is such that
\begin{equation}\label{n4}
    T\propto (2DS^{2D}-Q^2).
\end{equation}

The extremal \abh of this $d$-dimensional \RN solution corresponds to (Eqs. (13) and (14) of~\cite{hd})
\begin{equation}\label{n5}
Q^2=2DM^2,\;Q^2=2DS^{2D},\;T\equiv 0.
\end{equation}

According to Eqs. (8) and (34) of~\cite{hd}, the coefficient
\begin{equation}\label{q2}
    \mathcal{A}_3=(6d-14)\phi^2-(d-2)
\end{equation}
is proportional to $S(\partial H/\partial S)_{\phi}+\phi(\partial H/\partial \phi)_{S}$. Since the authors of~\cite{hd} assumed, in the paragraph following Eq. (37) of~\cite{hd}, that $S(\partial H/\partial S)_{\phi}+\phi(\partial H/\partial \phi)_{S}\propto H$, they concluded that the right-hand sides in~\eqref{q1} and~\eqref{q2} are proportional, which resulted in: $H=0\Leftrightarrow \mathcal{A}_3=0$ . First of all, this is not possible, since $H=0$ (or $\mathcal{B}_3=0$ and $S\neq 0$), results in $\phi^2=1/(2D)$ leading to $\mathcal{A}_3=2(d-1)/D\neq 0$. Second, $H$ as given in~\eqref{q1} is not homogeneous in ($S,\phi$) nor is it homogeneous in ($S^r,\phi^t$) for all $r\neq 0$ and $t\neq 0$, for it is not possible to find $r\neq 0$ and $t\neq 0$ such that $H(\la S^r,\la \phi^t)=\la^{\bt}H(S^r,\phi^t)$.

We see that $H=0$ ($\mathcal{B}_3=0$) leads to $\phi^2=1/(2D)$ or, using~\eqref{q1b}, to $S^{2D}=Q^2/(2D)$, which is the extremal black hole~\eqref{n5} where the temperature~\eqref{n4} vanishes but $\mathcal{A}_3\neq 0$. Thus, the conclusion drawn in the paragraph following Eq. (37) of~\cite{hd}, asserting that $g_H^{II}$ is singular, is not valid; rather, the metric $g_H^{II}$ (Eq. (34) of~\cite{hd}) is not singular or degenerate in the extremal black hole limit since $\det g_H^{II}\neq 0$.

We conclude that the scalar curvature diverges for $H=0$ (Eq. (35) of~\cite{hd}) while the metric $g_H^{II}$ remains regular. This should signal, according to GTD itself (see paragraph following Eq. (6) of~\cite{hd}), a second order phase transition while the thermodynamic classical description asserts no phase transition in this case (see paragraph following Eq. (21) of~\cite{hd}). This discrepancy (1) constitutes a failure to describe the case $\Phi=H$ by GTD or (2) may lead to modify the form of the metric $g^{II}$ in Eq. (8) of~\cite{hd}. One should also question the thermodynamic classical treatment performed in~\cite{hd} in the case $\Phi=H$. However, we verify that the discrepancy persists.

\vspace{3mm}

\paragraph*{\textbf{(b) Charged and rotating \BH.}} Another instance of misleading result in GTD occurred in the paragraph following Eq. (13) of~\cite{gbh} where the misleading equation $\bt M=TS+\Om_HJ+\phi Q$ was used to justify the presence of the factor $M$ in Eq. (11) of~\cite{gbh}. By writing this, the authors have thus assumed that all $p_a$'s are equal without, however, fixing the value of $\bt$.

The correct equation is $M/2=TS+\Om_HJ+\phi Q/2$ [see Eqs. (2.6) to (2.9) of~\cite{d2}], thus the conformal factor present in Eq. (11) of~\cite{gbh}, $TS+\Om_HJ+\phi Q$, is rather proportional to $M+\phi Q$ and not to $M$.

As is clear from the two previous examples, the authors of GTD have always treated equally the natural extensive variables ($E^a$) expressing the first law and the modified extensive variables ($E^{\,\prime a}$) in which the potential is homogeneous: Whenever they deal with a thermodynamic potential of some number of variables, $f(x,y,z,\ldots)$, they write $\bt f=x\partial f/\partial x+y\partial f/\partial y+\cdots$ or $f\propto x\partial f/\partial x+y\partial f/\partial y+\cdots$ even if $f$ is not homogeneous as in~\eqref{q1}. In \abh thermodynamics, the shape of Euler identity, which is not fixed a priori, is determined only once the explicit mathematical expression of $f(x,y,z,\ldots)$ is known.

\vspace{3mm}

\paragraph*{\textbf{(c) Kerr \bh in $d$-dimensions.}} A final point in our comments is the following, rather interesting, example.

First consider Eq. (47) of~\cite{hd} (Kerr \abh in $d$-dimensions):
\begin{equation}\label{i0}
  g_S^{II}=-\frac{M-\Om J}{T^2(TS+\Om J)}\,g_M^{II}.
\end{equation}
This last equation is a straightforward application of the change of representation formula, Eq. (53) of~\cite{conf}, which was derived by the authors of GTD taking $\bt =1$ and all $p_a\equiv 1$ [see Eq. (34) of~\cite{conf}]:
\begin{equation}\label{i1}
  g^{E^{(i)}}=-\Big[I_{(i)}^{-1}E^{(i)}\frac{1}{I_aE^a}\Big]g^{\Phi} \quad [\Si \text{ over }a,\text{ no }\Si \text{ over }(i)].
\end{equation}
We stress that the realm of applicability of the change of representation formula~\eqref{i1} is restricted by the constraints $\bt =1$ and all $p_a\equiv 1$ the authors have imposed. For instance, Eq.~\eqref{i1} does not apply to cases where all $p_a$'s are equal but all different from 1. As shown in the Appendix, this is precisely the case of Kerr \bh in $d$-dimensions.

Now back to Kerr \bh in $d$-dimensions. The authors of~\cite{hd} obtained~\eqref{i0} from~\eqref{i1} on substituting: $E^{(i)}=S$, $\Phi=M$, $I_{(i)}=T$, $I_aE^a=TS+\Om J$, $g^{E^{(i)}}=g_S^{II}$ and $g^{\Phi}=g_M^{II}$. This is an inappropriate application of~\eqref{i1} since the authors did not check whether all $\bar{p}_a$ are equal to 1. To show that explicitly, note that the direct substitution of $E^{(i)}=S$, $\Phi=M$, $I_{(i)}=T$, $I_aE^a=TS+\Om J$, $g^{E^{(i)}}=g_S^{II}$ and $g^{\Phi}=g_M^{II}$ in~\eqref{i1} yields the same expression as~\eqref{i0} but with $ST$ in the numerator instead of $M-\Om J$:
\begin{equation}\label{i0b}
  g_S^{II}=-\frac{ST}{T^2(TS+\Om J)}\,g_M^{II}.
\end{equation}
To reduce~\eqref{i0b} to~\eqref{i0}, the authors have assumed $M(S,J)=TS+\Om J\;[=(\partial M/\partial S)S+(\partial M/\partial J)J]$ thus taking $\bt=1$ and all $p_a\equiv 1$ for Kerr \bh in $d$-dimensions. Where does such a formula, $M(S,J)=TS+\Om J$, come\footnote{And where does the formula $U(S,V)=ST-PV$, which has been used in Eq. (20) of~\cite{nd}, come from? Here $U(S,V)$ is supposed to be arbitrary in~\cite{nd}, thus not known explicitly. Such formula is not even valid for a monatomic ideal gas with $PV=nRT$ and $U=3nRT/2$ for this would lead to $S=$ constant.} from? According to the second paragraph following Eq.~\eqref{note} and \cite{conf2}, we can always choose $\bt =1$ but once this is done, as we shall see also in the Appendix, all $p_a$ acquire well fixed values [Eqs.~\eqref{1b}, \eqref{A3}] that are functions of the parameters of the problem.

Moreover, it is straightforward to check that $M(S,J)=TS+\Om J$ is not correct by evaluating its right-hand side using the expressions of $T(=\partial M/\partial S)$ and $\Om (=\partial M/\partial J)$ given in Eq. (42) of~\cite{hd}, then comparing the result with the expression of $M$ given in Eq. (41) of~\cite{hd}. Rather, the correct expression is (see Appendix):
\begin{equation}\label{i2}
    DM(S,J)=TS+\Om J
\end{equation}
which reduces to Eq. (2.9) of~\cite{d2} (with $Q=0$) if $d=4$ [$\Rightarrow D=1/2$ by~\eqref{g0}].

As shown in the Appendix, and is obvious from~\eqref{i2}, $M(S,J)$ is homogeneous in ($S^D,J^D$) of order 1 or homogeneous in ($S,J$) of order $D$. We will work with the former option. But, with $\bt =1$, $p_1=p_S=D\neq 1$ and $p_2=p_J=D\neq 1$, so we cannot use~\eqref{i1}, which was derived assuming $\bt =1$ and all $p_a\equiv 1$ [see Eq. (34) of~\cite{conf}].

We first had to generalize~\eqref{i1} to include the case where $p_a\neq 1$~\cite{conf2}. Thus, if the order of homogeneity is chosen equal 1 and all or some $\bar{p}_a\neq 1$, then~\cite{conf2}
\begin{multline}\label{i3}
\hspace{-2.3mm}g^{E^{(i)}}=-\frac{\Phi-\sum_{j\neq i}I_jE^j+\sum_{j\neq i}(\bar{p}_{(i)}^{\ -1}-\bar{p}_j^{\ -1})I_jE^j}{I_{(i)}^2(I_aE^a)}\,g^{\Phi}, \\ [\Si \text{ over }a,\;(i)\text{ fixed}].
\end{multline}
where $\bar{p}_a$ are the values of $p_a$'s corresponding to an order of homogeneity equal 1 [see Eq.~\eqref{3d}], $\sum_{j\neq i}I_jE^j=I_aE^a-I_{(i)}E^{(i)}$ and $\Phi$ is given by~\eqref{3c} [or by~\eqref{3} on setting $\bt=1$ and $p_a=\bar{p}_a$]: $\Phi=I_{(i)}E^{(i)}/\bar{p}_{(i)}+\sum_{j\neq i}I_jE^j/\bar{p}_j$.

Applying~\eqref{i3} to Kerr \bh in $d$-dimensions with
\begin{align*}
&\text{all } \bar{p}_a=D,\quad (\bar{p}_{(i)}=D, \;\bar{p}_j=D),\\
&E^{(i)}=S,\; I_{(i)}=T,\; I_aE^a=TS+\Om J,\; \sum_{j\neq i}I_jE^j=\Om J,\\
&\Phi=(TS+\Om J)/D=M\; [\text{see Eq.~\eqref{i2}}],\\
&g^{E^{(i)}}=g_S^{II},\; g^{\Phi}=g_M^{II},
\end{align*}
we obtain
\begin{equation}\label{c1}
  g_S^{II}=-\frac{M-\Om J}{T^2(TS+\Om J)}\,g_M^{II},
\end{equation}
which is Eq.~\eqref{i0} of this paper [Eq. (47) of~\cite{hd}] that the authors of~\cite{hd} have reached upon using inappropriately formula~\eqref{i1} and admitting that $M(S,J)=TS+\Om J$ holds for Kerr \bh in $d$-dimensions.

The fact that the authors of~\cite{hd} have reached the correct formula~\eqref{c1} is due, as explained in the Conclusion section, to the property that all $\bar{p}_a$ are equal. This property makes  the conformal factor, $I_aE^a=TS+\Om J$, that the authors have chosen, proportion to $\Phi=M$, as Eq.~\eqref{i2} shows.

The case where all $p_a$ (or $\bar{p}_a$) are equal is not always met (see Appendix). Even if all $p_a$ are equal but different from 1, formula~\eqref{i1} is still not valid. From this point of view, Eq. (54) of~\cite{conf} and Eq. (20) of~\cite{nd}, where~\eqref{i1} has been used, are not valid because $U(S,V)$ is not known explicitly to assert that all $p_a\equiv 1$. In these last two references, the authors, applying inappropriately formula~\eqref{i1}, thought of $ST$ as $U+PV$, thus they assumed $U(S,V)=ST-PV$ to be a universal law, that is, $U(S,V)$ is homogeneous in ($S,V$) of order 1 for all thermodynamic systems. But such law does not even apply to an ideal gas where we have: $U=ST-PV+\mu N$ with $N$ being the one-component particle number, $\mu=-kT\ln(AkT/P)$ is the chemical potential, $A\equiv (2\pi mkT/h^2)^{3/2}$, $S=Nk\ln(A\e^{5/2}V/N)$, $U=3NkT/2$, and $kT/P=V/N$~\cite{book}.

Hence, for a general potential $U(S,V)$, the conclusion drawn in the paragraph following Eq. (21) of~\cite{nd} may no longer apply since the coefficient in Eq. (21) of~\cite{nd} has a more complicated structure, which is given by Eq.~\eqref{i3} of the present paper. This means that, besides the ambiguities that may occur if one uses $g_U^{II}$, as clarified in  the paragraph preceding section 4 of~\cite{nd}, other  ambiguities may take place if one uses $g_S^{II}$.

\section{Conclusion \label{sec4}}

We have concluded that the natural extrinsic \tv expressing the first law of thermodynamics are not the same variables in which thermodynamic potentials are homogeneous. This makes \abh thermodynamics a bit different from classical one. Generalizations of classical thermodynamics laws to apply to \abh thermodynamics are, however, possible and as example we derived the generalized Gibbs-Duhem relation and we extended the Euler identity. Other generalizations were made in~\cite{conf2}.

The misleading results and conclusions by the authors of GTD, due the indiscriminate use of the natural \tv and modified ones in \abh thermodynamics, has lead us to discover and reveal some other ambiguities and inconsistencies in the theory which were never discussed in the literature:
\begin{enumerate}
  \item The notion of ensembles is ambiguous in GTD;
  \item How is the conformal factor, which appears in the metric of GTD and is usually taken as $E^a\partial\Phi/\partial E^a$ ($\Si$ over $a$), related to ensembles? Is there a one-to-one relationship from the set of conformal factors to the set of ensembles? If not, and mostly this is going to be the case, there should be an equivalent relation regrouping different conformal factors into equivalent sets where a representant from each set is in a one-to-one relation with an element from the set of ensembles;
  \item It might seem possible to solve some inconsistencies in GTD had we chosen this conformal factor proportional to $\Phi$, that is, of the form $(E^a/\bar{p}_a)\partial\Phi/\partial E^a$ ($\Si$ over $a$) if $\Phi$ were homogeneous. This is true for the case (c) of Sect.~\ref{sec3} where no inconsistency occurs since the authors of~\cite{hd} have taken the conformal factor $=$ $E^a\partial\Phi/\partial E^a\propto (E^a/\bar{p}_a)\partial\Phi/\partial E^a$, which results from the fact that all $\bar{p}_a$'s are equal.

      However, if the conformal factor is different from \\$E^a\partial\Phi/\partial E^a$, one needs to modify the change of representation formula~\eqref{i3}. If this factor is taken equal to $\Phi$, we replace $I_aE^a$ in the denominator of~\eqref{i3} by $\Phi$, so the equation becomes:
      \begin{equation}
      \frac{g^{E^{(i)}}}{g^{\Phi}}=-\frac{\Phi-\sum_{j\neq i}I_jE^j+\sum_{j\neq i}(\bar{p}_{(i)}^{\ -1}-\bar{p}_j^{\ -1})I_jE^j}{I_{(i)}^2\Phi}.
      \end{equation}
      Other successful choices of this factor were made by the authors of GTD~\cite{conf}, among which we find the form $\xi^a_{b} I_aE^b$. In spite of what has been done in this work, this latter choice may not be one of the appropriate choices for \abh thermodynamics since it makes use of natural extensive \tv instead of the modifed ones. A more appropriate choice could be $\xi^a_{b} I_aE^b/p_b$. If this is the case, one needs to replace the factor $\xi^a_{b} I_aE^b$ in Eq. (20) of~\cite{conf2} by $\xi^a_{b} I_aE^b/p_b$, yielding
      \begin{multline}
      g^{E^{(i)}} = -\frac{1}{\beta I_{(i)}}  \bigg[ \frac{\xi^{(i)}_{(i)} E^{(i)}}{p_{(i)}}  +  \sum_{j\neq i}\bigg(\frac{\xi^{(i)}_{(i)}}{p_{(i)}} - \xi^j_{j}\beta \bigg) \frac{I_jE^j}{I_{(i)}}\bigg]\times\\ \frac{g^{\Phi}}{(\xi^a_{b} I_aE^b/p_b)}.
      \end{multline}
  \item If $\Phi$ is not homogeneous, as in the case (a) of Sect.~\ref{sec3}, one may think to define this conformal factor using generalized homogeneous functions~\cite{art,book2}.

      Generalized homogeneous functions seem to be the most appropriate offered way to define the conformal factor even if $\Phi$ were homogeneous. In fact, these functions introduced for the first time in~\cite{art} have the properties that their derivatives and their Legendre transforms are also generalized homogeneous functions. The latter property is not satisfied in the change of representation in GTD made in~\cite[Sect. IV]{conf} where it is admitted that the new representation $E^{(i)}$ is not a homogeneous function when the old representation $\Phi$ is.
\end{enumerate}

\section*{\large Appendix: Smarr formula for Kerr \abh in $d$-dimensions}
\renewcommand{\theequation}{A.\arabic{equation}}
\setcounter{equation}{0}

The purpose is to show that $M(S,J)$ as given by Eq. (41) of~\cite{hd}
\begin{equation}\label{A1}
    M(S,J)=\frac{d-2}{4}S^D\Big[1+\frac{4J^2}{S^2}\Big]^{1/(d-2)}
\end{equation}
is homogeneous in ($S^D,J^D$) of order 1 [or, equivalently, homogeneous in ($S,J$) of order $D$]. Assume that \\$M(\la S^{p_S},\la J^{p_J})=\la^{\bt}M(S^{p_S},J^{p_J})$. To determine $p_S$, $p_J$ in terms of $\bt$ we evaluate the right-hand side of~\eqref{A1} at the point ($\la^{1/p_S}S,\la^{1/p_J}J$)
\begin{equation}\label{A2}
    \frac{d-2}{4}\la^{D/p_S}S^D\Big[1+\frac{4\la^{2/p_J}J^2}{\la^{2/p_S}S^2}\Big]^{1/(d-2)}
\end{equation}
which we set equal to $\la^{\bt}M(S,J)$. This leads to
\begin{multline}\label{A3}
p_S=p_J\, \text{ and }\,\bt=D/p_S\\ \text{ or: } p_S(\bt)=p_J(\bt)=D/\bt.
\end{multline}
This is a special case where all $p_a$ are equal. If, in~\eqref{A3}, we choose $\bt=1$, we obtain $p_S=p_J=D$ and we are led to
\begin{equation}\label{A4}
    M(\la S^D,\la J^D)=\la M(S^D,J^D)
\end{equation}
where $M(S^D,J^D)$ is not the value of the right-hand side of~\eqref{A1} evaluated at the point ($S^D,J^D$); rather it is the same expression~\eqref{A1} with ($s,j$) $=$ ($S^D,J^D$) taken as independent variables:
\begin{equation}\label{A5}
    M(S^D,J^D)=\frac{d-2}{4}S^D\Big[1+\frac{4(J^D)^{2/D}}{(S^D)^{2/D}}\Big]^{1/(d-2)}.
\end{equation}
If we choose $\bt=D$, we obtain $p_S=p_J=1$ and we are led to
\begin{equation}\label{A6}
    M(\la S,\la J)=\la^D M(S,J).
\end{equation}
Both equations~\eqref{A4} and~\eqref{A6} are correct and lead to the same Euler identity~\eqref{i2}, which can be verified on evaluating its right-hand side using the expressions of $T=\partial M/\partial S$ and $\Om=\partial M/\partial J$ given in Eq. (42) of~\cite{hd}. This also confirms the fact that the $p_a$'s depend on $\bt$ but the product $\bt p_a(\bt)=\bar{p}_a$ does not [Eq.~\eqref{3d}].

If we consider the thermodynamics of \RN \bh in $d$-dimensions~\cite{RN} and apply the same procedure to Eq. (12) of~\cite{hd} assuming $M(\la S^{p_S},\la Q^{p_Q})$ $=$\\ $\la^{\bt}M(S^{p_S},Q^{p_Q})$ we find $p_S(\bt)=D/\bt$, $p_Q(\bt)=1/\bt$~\cite{conf2}. If we choose $\bt=1$, this leads to $p_Q=1$, $p_S=D$. On applying~\eqref{3} we obtain $DM=TS+D\phi Q$ with $T=(\partial M/\partial S)_Q$, $\phi=(\partial M/\partial Q)_S$~\cite{conf2}. In this case, it is not possible to have $p_Q=p_S$ for all $\bt$.

%\newpage

\end{document}